\documentclass[aps,prl,twocolumn,letterpaper]{revtex4}
\usepackage{amsmath,amssymb,amsfonts,amsthm}
\usepackage{graphicx,bm}

\begin{document}

\title{Low-temperature phase transformations of PbZr$_{1-x}$Ti$_{x}$O$_{3}$
in the morphotropic phase-boundary region}
\date{}
\author{F. Cordero,$^{1}$ F. Craciun,$^{1}$ and C. Galassi$^{2}$}
\affiliation{$^1$ CNR-ISC, Istituto dei Sistemi Complessi, Area della Ricerca di Roma -
Tor Vergata,\\
Via del Fosso del Cavaliere 100, I-00133 Roma, Italy}
\affiliation{$^{2}$ CNR-ISTEC, Istituto di Scienza e Tecnologia dei Materiali Ceramici,
Via Granarolo 64, I-48018 Faenza, Italy}

\begin{abstract}
We present anelastic and dielectric spectroscopy measurements of PbZr$_{1-x}$%
Ti$_{x}$O$_{3}$ with $0.455\leq x\leq 0.53$, which provide new information
on the low temperature phase transitions. The tetragonal-to-monoclinic
transformation is first-order for $x<0.48$ and causes a softening of the
polycrystal Young's modulus whose amplitude may exceed the one at the
cubic-to-tetragonal transformation; this is explainable in terms of linear
coupling between shear strain components and tilting angle of polarization
in the monoclinic phase. The transition involving rotations of the octahedra
below 200~K is visible both in the dielectric and anelastic losses, and it
extends within the tetragonal phase, as predicted by recent first-principle
calculations.
\end{abstract}


\maketitle

Lead-titanate zirconate PbZr$_{1-x}$Ti$_{x}$O$_{3}$ (PZT) is a
ferroelectric perovskite widely used in applications, especially due
to its large piezoelectric response, and its phase diagram is known
since several decades \cite{JCJ71}. Yet, there is a resurgence of
interest for this material, especially focused on the so-called
morphotropic phase boundary (MPB) region, $x\simeq 0.44-0.52$,
separating the Ti-rich tetragonal (T with space group $P4mm$) phase
from the rhombohedral one (R = $R3m$). The numerous recent neutron,
X-ray and electron diffraction studies have been reviewed in Refs.
\onlinecite{GTB04,NC06}, but still new features are appearing
\cite{KBJ06}. The two ferroelectric phases are connected by an
intermediate monoclinic (M = $Cm$) phase, where the direction of the
polarization vector is intermediate between the T and R\ structures;
in addition, below $\sim 200 $~K a doubling of the monoclinic cell
has been reported \cite{RSR05}, which seems analogous to the
transition from high- to low-$T$ rhombohedral structure (from $R3m$
to $R3c$) at lower $x$, with antiphase rotations of the O octahedra
about the $\left[ 111\right] $ pseudocubic axis.

On the theoretical side, Haun \cite{HFJ89} proposed a comprehensive
description of the PZT phase diagram known at that time, without M
phase, based on the Landau expansion of the free energy in powers of
the polarization up to the sixth order \cite{HFJ89}; after the
discovery of the M phase, attempts have been made to obtain such a
phase within the same scheme, by introducing an additional order
parameter \cite{SLA00}, an additional tetragonal invariant
\cite{Hud06}, or terms beyond the sixth order \cite{VC01}. Also
microscopic descriptions have been provided, based on first
principle calculations \cite{BGV00,FS01}, and recently the inclusion
of antiferrodistortive degrees of freedom allowed the known phases
of PZT to be reproduced, plus an additional low-$T$ tetragonal
variant, again involving rotations of the octahedra \cite{KBJ06}.

We present anelastic and dielectric spectroscopy measurements of PZT with $%
0.455\leq x\leq 0.53$, providing new information on the nature of
the T-M transition and confirming the existence of a further
low-temperature transformation in the M phase and even in the T one
beyond the MPB region \cite{KBJ06}.

The ceramic samples have been prepared by the mixed-oxide method; the
starting oxide powders were calcined at 800~$^{\mathrm{o}}$C for 4 hours,
pressed into bars and sintered at 1200~$^{\mathrm{o}}$C for 2~h, packed with
PbZrO$_{3}$\ + 5wt\% excess ZrO$_{2}$\ in order to maintain a constant PbO
activity at the sintering temperature. The absence of impurity phases was
checked by powder X-ray diffraction and densities were about 96\% of the
theoretical ones. The ingots were cut into bars $4-5~$cm long and $0.5-0.6$%
~mm thick. Electrodes for the anelastic and dielectric spectroscopy
measurements were applied with silver paint and the samples were annealed in
air at 750~$^{\mathrm{o}}$C for avoiding any effects from the possibly
damaged surfaces after cutting. The dielectric susceptibility $\chi =\chi
^{\prime }-i\chi ^{\prime \prime }$ was measured with a HP 4194 A impedance
bridge with a four wire probe and a signal level of 0.5 V/mm, between 200~Hz
and 200~kHz. The measurements were made on heating/cooling at $1-1.5$~K/min
between 100 to 530~K in a Delta climatic chamber. The dynamic Young's
modulus $E\left( \omega ,T\right) =E^{\prime }+iE^{\prime \prime }$ or its
reciprocal, the elastic compliance $s=s^{\prime }-is^{\prime \prime }=E^{-1}$%
, was measured between 50 and 750~K by electrostatically exciting the
flexural modes of the bars suspended in vacuum on thin thermocouple wires in
correspondence with the nodal lines. The real part of the Young's modulus is
related to the fundamental resonance frequency by $\omega /2\pi =1.028\frac{h%
}{l^{2}}\sqrt{\frac{E^{\prime }}{\rho }}$, where $h$ and $l$ are
sample thickness and length and $\rho $ its mass density
\cite{NB72}; these quantities depend on temperature much less than
$E$, so that in the following we report $s\left( T\right)
/s_{0}\simeq $ $\omega ^{2}\left( T_{0}\right) /\omega ^{2}\left(
T\right) $ with $T_{0}=750$~K. Also the
third and fifth modes could be measured, whose frequencies are $5.4$ and $%
13.3$ times greater; the fundamental frequencies of the various samples
where $\omega /2\pi \simeq 1-1.6$~kHz, and the Young's moduli at high
temperature, uncorrected for porosity, were $E_{0}=s_{0}^{-1}=$ 138, 139,
134 and 125~GPa for $x=0.455$, 0.465, 0.48 and 53, respectively.

\begin{figure}
   \includegraphics[width=8.5 cm]{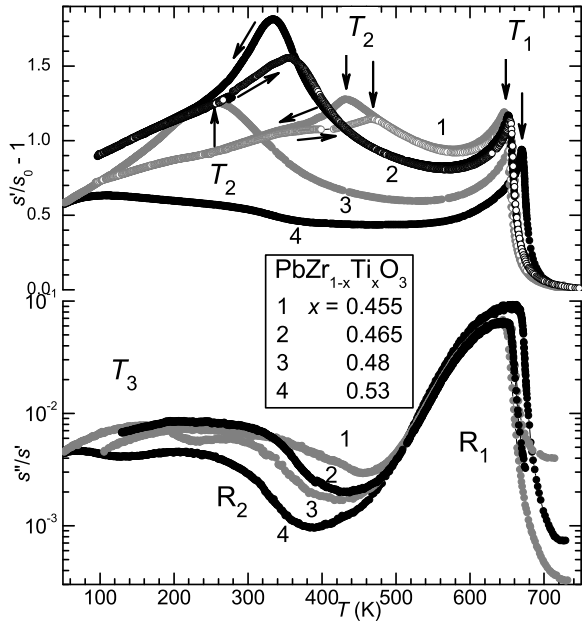}\\
  \caption{Anelastic spectra of PbZr$_{1-x} $Ti$_{x}$O$_{3}$ with varying
  Ti/Zr ratio. In the real part, both the cooling/heating (filled/empty
  symbols) curves are shown; $f=1-1.6$~kHz at high temperature.}
  \label{fig an}
\end{figure}

Figure \ref{fig an} presents the dynamic compliance curves normalized to $%
s_{0}=s^{\prime }\left( 750\text{~K}\right) $ for all samples. It is
possible to identify without ambiguities the transition temperatures
$T_{1}$ between paraelectric cubic (C) and ferroelectric T phases,
and $T_{2}$ between T and M. Both produce maxima in $s^{\prime
}\left( T\right) $, whose temperatures coincide with the
corresponding boundary lines in the PZT\ phase diagram (see Fig.
\ref{fig pd} below); the identification of the maximum in $s^{\prime
}$ at $T_{2}$ with the T-M transformation has already been proposed
\cite{BBG05}. The C-T transition is almost second order, as known
for Zr-rich compositions, with small hysteresis between cooling and
heating, $\Delta T<3$~K, and is accompanied by a sharp rise of the losses $%
s^{\prime \prime }/s^{\prime }$ on approaching the T phase. The transition
at $T_{2}$ is clearly first order for $x=0.455$, with $\Delta T=35$ K, but $%
\Delta T$ decreases with increasing Ti content and is almost null at $x=0.48$%
. The maximum of $s^{\prime }$ at $T_{2}$ is not accompanied by any clear
anomaly in the losses, which however exhibit an increase in this temperature
region in all samples. Such a broad dissipation maximum, labeled R$_{2}$, is
present also at $x=0.53$, where no T-M transition occurs, and is found also
in the dielectric loss $\chi ^{\prime \prime }/\chi ^{\prime }$, so that we
suppose that it is mainly due to relaxational dynamics of domain walls (DW).
At lower temperature, there is no clear anomaly in $s^{\prime }$, while
those in $s^{\prime \prime }$ connected with the transition at $T_{3}$ will
be described below.

\begin{figure}
   \includegraphics[width=8.5 cm]{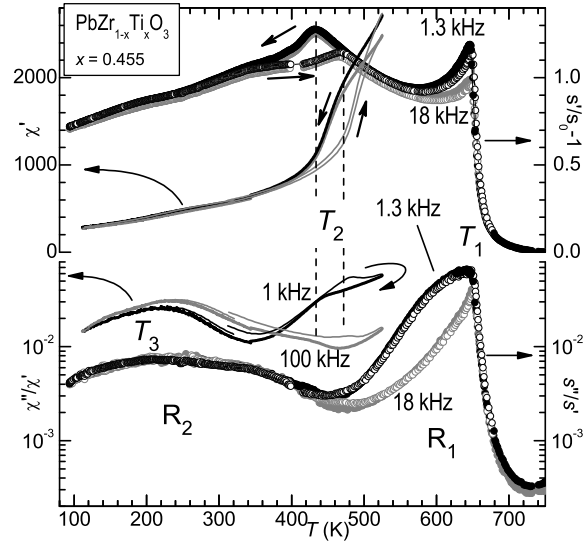}\\
  \caption{Dielectric and
anelastic spectra of PZT with $x=0.455$ measured at two frequencies
on cooling/heating (thick/thin lines and filled/empty symbols).}
  \label{fig Ti45}
\end{figure}

A\ comparison between anelastic and dielectric susceptibilities is presented
in Fig. \ref{fig Ti45} for $x=0.455$. The temperature hysteresis connected
with the T-M transformation is found also in the $\chi \left( \omega
,T\right) $ curves, but the maximum in $\chi ^{\prime }$ is much less
pronounced than in $s^{\prime }$, also because it is masked by the rise of $%
\chi ^{\prime }$ on approaching the ferroelectric transition at $T_{1}$.
Rather, a maximum can be identified in the dielectric loss, as evidenced by
the dotted vertical lines. The frequency dispersion at high temperature in $%
\chi \left( \omega ,T\right) $ should be due to the thermally
activated motion of mobile charged defects, presumably O vacancies
accompanying Pb vacancies formed during sintering, and/or charge
carriers from the ionization of such defects; the motion of such
charges between the DW under the applied ac field causes a
relaxation of the polarization (space charge relaxation
\cite{LG77}). Mobile charges may also produce anelastic relaxation
by migrating to the 90$^{\mathrm{o}}$ DW that become charged during
the sample vibration due to piezoelectric effect \cite{Gri90}; the
thermally activated (shifting to higher $T$ for higher $\omega $)
tail in $s^{\prime
\prime }/s^{\prime }$ disappearing with the DW above $T_{1}$ , labeled R$%
_{1} $, may be due to both this mechanism and to dissipative motion of DW.
The same relaxation is also responsible for the frequency dependent rise of $%
s^{\prime }\left( \omega ,T\right) $ on approaching $T_{1}$ from below, so
that the contribution of the C-T phase transformation is actually a step in $%
s^{\prime }$ plus a peaked contribution due to fluctuations responsible for
the gradual rise of $s$ on approaching $T_{1}$ from above.

The step-like behavior of $s^{\prime }$ at $T_{1}$ corresponds to the fact
that $s$ is the strain susceptibility, and strain $e$ is coupled to the
primary order parameter, the polarization $P$, by terms that are linear in $%
e $ and quadratic in $P$ (polarization reversal does not change
strain). According to the widely used Landau expansion of the free
energy in powers of the order parameter, such linear-quadratic
coupling yields a renormalization of the elastic constants which has
a step at the transition temperature, but no additional temperature
dependence \cite{Reh73,Sal90,II99b}.

We consider now the T-M transition at $T_{2}$, mainly signaled by a peak in $%
s^{\prime }$ and temperature hysteresis in all the susceptibilities.
The T-M phase transformation is predicted to be second order
\cite{VC01} or near-to-second order \cite{Hud06} and the crossover
to first-order character with decreasing $x$ may be understood by
considering that $x=0.455$ is at the boundary between R and M\
phases (Fig. \ref{fig pd}), and the T-R transition must be of the
first order, lacking a group-subgroup relationship between the T and
R space groups.

A most remarkable feature is the amplitude of the peak in $s^{\prime }\left(
T\right) $, which is even larger than the step at $T_{1}$ for $x=0.465$ and
0.48; in fact, as noted above for the C-T transition at $T_{1}$, coupling
terms between polarization $P$ and strain $e$ of the type predicted for PZT
are linear in $e$ and quadratic in $P$, and should therefore yield a step
and not a maximum in the elastic constants. The peak exhibited by $s\left(
\omega ,T\right) $ at $T_{2}$ is frequency independent (see Fig. \ref{fig
Ti45}) and therefore cannot be attributed to the motion of domain walls, as
for the rising part of $s\left( \omega ,T\right) $ below $T_{1}$; in
addition, it can hardly be attributed to fluctuations, which are not
expected to play a significant role for a first-order transition like that
at $x\simeq 0.46$ and should also appear in the imaginary part. The most
obvious explanation for such a large effect on $s^{\prime }$ is a linear
coupling between strain and order parameter; in that case, strain follows
the order parameter and a Curie-Weiss-like divergence is expected in the
corresponding elastic compliance, as in a proper ferroelastics \cite%
{Reh73,Sal90}. We show that this explanation is possible at a T-M
transition characterized by a deviation of the polarization from the
tetragonal $c$ axis with little change of the magnitude, namely
$P_{1}=$ $P_{2}$ gradually increase at $T_{2}$ with $P_{3}$
remaining almost constant, as supposed by Hudak \cite{Hud06} and
verified in Monte Carlo simulations \cite{KBJ06}. The term that
couples $T_{2g}$-type shears, $e_{i}$ with $i=4-6$ (Voigt notation:
$e_{4}=e_{23}$, etc.) with polarization is \cite{HFJ89,II99b,Hud06}
\begin{equation*}
F_{P-e^{\text{T}}}=Q_{44}\left(
e_{4}P_{2}P_{3}+e_{5}P_{3}P_{1}+e_{6}P_{1}P_{2}\right) ~;
\end{equation*}%
if at the M-T transition the order parameter is $P_{1}=P_{2}\ll $ $%
P_{3}\simeq $ const, then the first two terms become\ bilinear in $e_{4}$, $%
P_{2}$ and $e_{5}$, $P_{1}$, and the change of the corresponding elastic
compliances $s_{44}$ and $s_{55}$ is expected \cite{Reh73,Sal90} to be $%
\Delta s_{44}^{\text{M}}=\Delta s_{55}^{\text{M}}\propto \left(
T_{0}-T\right) ^{-1}$, where $T_{0}\simeq T_{1}$ is the temperature
at which the Landau coefficient of the term quadratic in $P$
vanishes. The shape of the peaks in $s$ even suggests that the
divergence occurs at $T_{0}\simeq T_{2}$, as if there was a new
order parameter associated with the transverse polarization $P_{1}$
or, almost equivalently for $P_{1}\ll P_{3}$, to the tilt angle of
the polarization away from the $c$ axis \cite{SLA00,Hud06}. Yet, we
are prevented from making a strong statement on this point from the
complications involved in the determination of the contributions of
these shear elastic constants to the Young's modulus of the
ferroelectric polycrystal, from the temperature dependence of other
elastic constants \cite{Hud06} and from the possible broadening due
to even small inhomogeneities in the Ti/Zr ratio, whose effect is
amplified by the steep slope of the $T_{2}\left( x\right) $ line (25
times steeper than $T_{1}\left( x\right) $, see Fig. \ref{fig pd}).

These measurements address the issues raised by Vanderbilt and Cohen
\cite{VC01} on the existence of a softening of a shear elastic
constant, as implied by treatments of the M phase in terms of free
energy expansions up to the sixth order only, and on temperature
hysteresis versus compositional inhomogeneity for explaining the
coexistence of T and M phases observed in diffraction experiments
\cite{NCS00}. The softening of the shear elastic constants $c_{44}$
and $c_{55}$ indeed exists and is large at the T-M transition,
though we cannot tell with certainty whether the tilt angle of
polarization from the $c$ axis is a real order parameter requiring a
new instability \cite{SLA00,VC01}; the existence of such a softening
should anyway serve as a guide in constructing the ever improving
microscopic models of PZT and similar ferroelectric perovskites
\cite{BGV00,FS01,KBJ06}. Regarding the nature of the T-M
transformation, we confirm that it is second-order well within the M
phase, but becomes first-order with temperature hysteresis for
$x<0.48$, so that the region of coexistence between T and M observed
by Noheda (hatched in Fig. \ref{fig pd}) does not imply
compositional inhomogeneity of their samples.

\begin{figure}
   \includegraphics[width=8.5 cm]{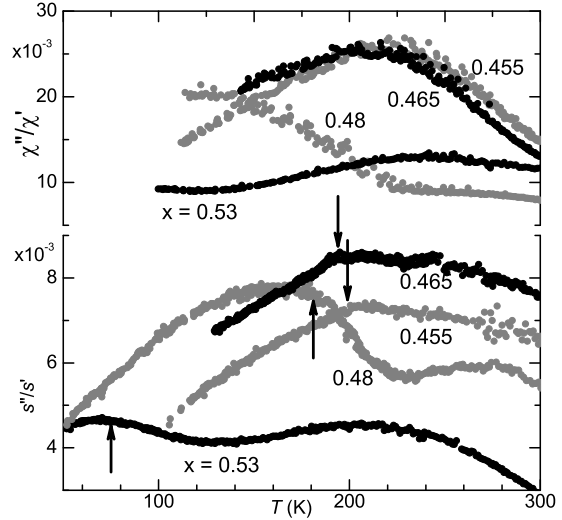}\\
  \caption{Dielectric and
anelastic spectra of PZT with $x=0.455$ measured at two frequencies
on cooling/heating (thick/thin lines and filled/empty symbols).}
  \label{fig Qtand}
\end{figure}

We discuss now the anomalies at $T_{3}$, connected with the possible
onset of octahedral rotations. An experimental $T_{3}$ line exists
only for the R phase, which passes from $R3m$ to $R3c$ with tilting
of the O\ octahedra about $\left[ 111\right] $ directions, while its
extension into the M\ phase is controversial
\cite{FIE02,RSR05,CNS05,NC06} and that in the T phase is only
theoretical \cite{KBJ06}. In fact, it is still controversial whether
superlattice reflections observed in diffraction spectra at liquid
He temperatures are due to rotations of the octahedra, with the M
phase passing from $Cm$ to $Cc$ \cite{RSR05}, or instead to
coexistence of the $Cm$ and $R3c$ structures
\cite{FIE02,CNS05,NC06}. Figure \ref{fig Qtand} shows that an
anomaly within the M phase indeed occurs in the imaginary parts of
the susceptibilities, while nothing noticeable is found in the real
parts. The arrows in Fig. \ref{fig Qtand} point to a transition
temperature $T_{3}$, under the assumption that the elastic losses
behave as in most cases at phase transitions, namely with a rise on
cooling down to the transition temperature, where a more or less
sharp edge is found, depending on frequency (see Fig. \ref{fig Ti45}
at $T_{1}$); a clear rise is not observed for $x=0.455$ and 0.465,
but it might be absorbed by a decrease of the relaxation R$_{2}$
(Fig. \ref{fig an}) and a sharp edge is identifiable. The dielectric
loss presents a broader step or peak, whose temperature correlates
with that in the $s^{\prime \prime }$ step; at the higher Ti
content, $\chi ^{\prime \prime }/\chi ^{\prime }$ starts changing
slope around 120~K, but the step could not be measured because of
limitations in the temperature range. The curves in Fig. \ref{fig
Qtand} are measured on cooling exciting the first flexural mode for
$s$ and at 1~kHz for $\chi $, but they are perfectly reproduced on
heating and an increase of frequency slightly increases the overall
dissipation level leaving unchanged the shape and temperature of the
anomalies; in addition, they are unaffected by annealing in air at
750~$^{\mathrm{o}}$C in order to eliminate possible surface damage
after cutting the samples. These features indicate that a structural
transformation indeed occurs at $T_{3}$ and is a bulk rather than a
surface effect, so confirming the extension of the $T_{3}$ line in
the M phase.

\begin{figure}
  \includegraphics[width=8.5 cm]{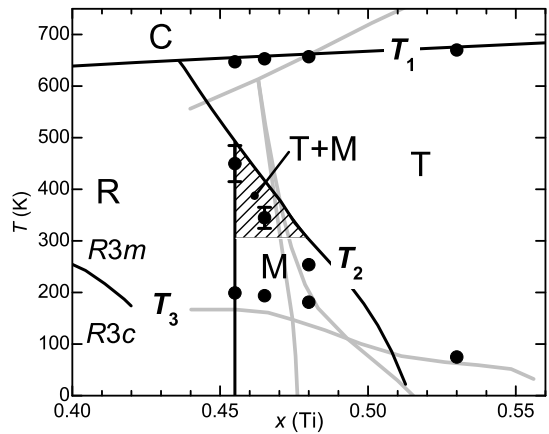}\\
  \caption{Experimental \protect\cite%
{JCJ71,NC06} and calculated \protect\cite{KBJ06} (gray lines) phase
diagram of PZT. The line $T_{3}$ is the onset of transitions
involving rotations of the O octahedra. The full symbols are the
transition temperatures deduced
from the anelastic spectra: $T_{1}$ and $T_{2}$ from the maxima in $%
s^{\prime }$ (the error bars indicate the hysteresis between cooling
and heating) and $T_{3}$ from the kink in $s^{\prime \prime }$.}
\label{fig pd}
\end{figure}

The appearance of a clear step in $s^{\prime \prime }$ even at 53\%
Ti, well within the T phase, further confirms the prediction of the
antiferrodistortive instability also in this phase \cite{KBJ06}. Figure \ref%
{fig pd} presents the established experimental phase diagram
\cite{JCJ71,NC06} together with the recently calculated one
\cite{KBJ06} (gray lines), and with the transition temperatures
deduced from the present study; there is good agreement between
experimental and theoretical $T_{3}\left( x\right) $ lines, and also
with the region of coexistence of M and T\ phases, where we observe
hysteresis between heating and cooling.

In conclusion, the low temperature phase transformations of PZT at
compositions near the MPB cause well detectable effects both in the
anelastic and dielectric responses. The T-M transition is shown to be first
order with temperature hysteresis at $x=0.455$, on the border between M and
R phases, and gradually becomes second order (without temperature
hysteresis) increasing $x$ into the M region. Such a transition is
accompanied by a large softening of the Young's modulus, that can be
explained in terms of Curie-Weiss-like behavior of the $s_{44}$ and $s_{55}$
compliances in the M phase, since the deviation angle of the polarization
from the tetragonal $c$ axis is linearly coupled with the $e_{4}$ and $e_{5}$
strains. The low temperature transformation involving antiferrodistortive
tilting of the octahedra is confirmed to exist in the M and even in the T
phase, at least up to a Ti fraction of 53\%.

We thank O. Hudak for useful discussions and F. Corvasce, M. Latino and C.
Capiani for technical assistance.

%\bibliographystyle{apsrev}
%\bibliography{biblio}

\begin{references}

\bibitem{JCJ71}
B. Jaffe, W. R. Cook and H. Jaffe, {\it Piezoelectric Ceramics}.  (Academic Press, London,
1971).

\bibitem{GTB04}
A.M. Glazer {\it et al.}, Phys. Rev. B {\bf 70}, 184123 (2004).

\bibitem{NC06}
B. Noheda {\it et al.}, Phase Transitions {\bf 79}, 5 (2006).

\bibitem{KBJ06}
I.A. Kornev {\it et al.}, Phys. Rev. Lett. {\bf 97}, 157601 (2006).

\bibitem{RSR05}
R. Ranjan {\it et al.}, Phys. Rev. B {\bf 71}, 92101 (2005).

\bibitem{HFJ89}
M.J. Haun {\it et al.}, Ferroelectrics {\bf 99}, 13 (1989).

\bibitem{SLA00}
A.G. Souza Filho {\it et al.}, Phys. Rev. B {\bf 61}, 14283 (2000).

\bibitem{Hud06}
O. Hudak, cond-mat/0609226 (2006).

\bibitem{VC01}
D. Vanderbilt {\it et al.}, Phys. Rev. B {\bf 63}, 94108 (2001).

\bibitem{BGV00}
L. Bellaiche {\it et al.}, Phys. Rev. Lett. {\bf 84}, 5427 (2000).

\bibitem{FS01}
M. Fornari {\it et al.}, Phys. Rev. B {\bf 63}, 92101 (2001).

\bibitem{NB72}
A.S. Nowick and B.S. Berry, {\it Anelastic Relaxation in Crystalline Solids}.
(Academic Press, New York, 1972).

\bibitem{BBG05}
A. Bouzid {\it et al.}, J. Eur. Ceram. Soc. {\bf 25}, 3213 (2005).

\bibitem{LG77}
M.E. Lines and A.M. Glass, {\it Ferroelectricity}.  (Oxford University Press,
Oxford, 1977).

\bibitem{Gri90}
S.A. Gridnev, Ferroelectrics {\bf 112}, 107 (1990).

\bibitem{Reh73}
W. Rehwald, Adv. Phys. {\bf 22}, 721 (1973).

\bibitem{Sal90}
E.K.H. Salje, {\it Phase transitions in ferroelastic and co-elastic crystals}.
(Cambridge University Press, Cambridge, 1990).

\bibitem{II99b}
Y. Ishibashi {\it et al.}, Jpn. J. Appl. Phys. {\bf 38}, 1454 (1999).

\bibitem{NCS00}
B. Noheda {\it et al.}, Phys. Rev. B {\bf 63}, 14103 (2000).

\bibitem{FIE02}
J. Frantti {\it et al.}, Phys. Rev. B {\bf 66}, 64108 (2002).

\bibitem{CNS05}
D.E. Cox {\it et al.}, Phys. Rev. B {\bf 71}, 134110 (2005).

\end{references}

\end{document}